# Nanostructured Hyperbolic Meta-Antennas Enable Arbitrary Control of Scattering vs Absorption

Nicolò Maccaferri[1,¶], Yingqi Zhao[1,¶], Marzia Iarossi[1,2], Tommi Isoniemi[1], Antonietta Parracino[1], Giuseppe Strangi[1,3,4], and Francesco De Angelis[1,*]

[1]*Istituto Italiano di Tecnologia, Via Morego 30, 16163, Genova, Italy*
[2]*DIBRIS, Università degli Studi di Genova, Via Balbi 5, 16126 Genova, Italy*
[3]*Department of Physics, Case Western Reserve University, 10600 Euclid Avenue, Cleveland, Ohio 44106, USA*
[4]*CNR-NANOTEC and Department of Physics, University of Calabria, 87036, Italy*

We show that meta-antennas made of a composite material displaying type II hyperbolic dispersion enable precise and controlled spectral separation of absorption and scattering processes in the visible/near-infrared frequency range. The experimental evidence is supported by a comprehensive theoretical study. We demonstrate that the physical mechanism responsible for the aforementioned effect lies in the different natures of the plasmonic modes excited within the hyperbolic meta-antennas. We prove that it is possible to have a pure scattering channel if an electric dipolar mode is induced, while a pure absorption process can be obtained if a magnetic dipole is excited. Also, by varying the geometry of the system, the relative weight of scattering and absorption can be tuned, thus enabling an arbitrary control of the decay channels. Importantly, both modes can be efficiently excited by direct coupling with the far-field radiation, even when the radiative channel (scattering) is almost totally suppressed, hence making the proposed architecture suitable for practical applications.

## I. INTRODUCTION

Unlike conventional optics, plasmonics enables unrivalled concentration of optical energy well beyond the diffraction limit of light, leading to extremely confined and enhanced electromagnetic fields at the nanoscale [1-7]. Besides its fundamental importance, manipulation of light at the subwavelength level is of great interest for the prospect of real-life applications [8], such as energy harvesting and photovoltaics [9-11], wave-guiding and lasing [12], optoelectronics [13], and biomedicine [14,15]. Along with the ongoing efforts to synthesize novel plasmonic materials to improve the performances of the aforementioned uses [16-18], novel optical designs and architectures that modify the optical power flow through plasmonic nanostructures represent another crucial step toward nanoscale manipulation of light-matter interactions [19]. Plasmonic nanostructures are known to exhibit, when coupled to light, collective electronic oscillations, so-called localized surface plasmon resonances (LSPRs), which determine their optical response in the visible and near-infrared spectral range.

*Corresponding author.
francesco.deangelis@iit.it
¶Contributed equally.

One of the drawbacks of plasmonic nanostructures exhibiting LSPRs is the spectral overlapping of scattering and absorption processes due to the intrinsic nature of the excited plasmonic mode, which is actually related to the optical properties of the constituent material. For guiding light, for instance, it is essential that the photonic or plasmonic circuit does not have a high absorption, while for other kind of applications, such as photo-acoustic imaging, it is crucial that the light is absorbed rather than scattered. To overcome these issues, one can shift the LSPR of interest, just modifying the geometry of the nanostructure, to reduce or increase the weight of the absorption compared to the scattering, although these two channels are at the same wavelength and one can choose to have only either absorption or scattering at a same time. An ideal solution would be an architecture and/or material which allows in the same platform a full control of the spectral distribution of scattering and absorption processes. In this framework, hyperbolic metamaterials (HMMs) [20-22] have received great attention from the scientific community due to their unusual and unexpected properties at optical frequencies, in particular in the near-infrared where, for instance, they can absorb more than 90% of the incident light [23,24]. These materials show rare properties never observed in nature [25-27], such as negative refraction [28-31] and resonant gain singularities [32], and can have a huge impact on nanoscale light confinement [33], optical cloaking [34], biosensing [35,36], nonlinear optics [37], super resolution imaging and superlensing effects [38], ultra-compact optical quantum circuits [39], plasmonic-based lasing [40], highly efficient artificial optical magnetism [41], graphene-based technologies [42], etc. When considering the dielectric tensor, HMMs can be divided into two types: type I has one negative component in its permittivity tensor and two positive ones. In contrast, a type II HMM has two negative components and one positive. In practical terms, type II appears as a metal in one plane and as a dielectric in the perpendicular axis, while type I is the opposite. Such anisotropic materials can sustain propagating modes with very large wave vectors and longer lifetime and propagation length in comparison to classic plasmonic materials [43] and exhibit diverging density of states [44], leading to a strong Purcell enhancement of spontaneous radiation [45-47]. Beyond the so-called natural hyperbolic materials, it is possible to mimic hyperbolic properties, for instance of type II, using a periodic stack of metallic and dielectric layers [48] that can support surface plasmons with large wave vectors [49] and whose effective permittivities for different polarizations have different signs [22,50].

In this work, we introduce a novel optical functionality of HMMs, focusing on an archetypical nanostructure, namely a cylindrically-shaped nanoantenna, made of artificial HMM of type II composed by alternating layers of metal and a dielectric material with a refractive index (RI) between 1.45 (such as $SiO_2$) and 2.25 (such as $TiO_2$). The proposed hyperbolic meta-antennas enable the creation of well-separated bands of either almost pure absorption or scattering and allow a full control of the ratio between these two channels over a broad spectral range in the visible and near-infrared regions. We provide a detailed study, supported by experimental evidence, where we explain the physical mechanism underlying the aforementioned effect of spectral separation between absorption and scattering processes. We show that the scattering band depends on the excitation of an electric dipolar mode, while the strong absorption band is due to the excitation of a magnetic dipolar mode. Interestingly, both modes can be excited by direct coupling with the external radiation, even when the scattering channel is practically suppressed, thus making the proposed architecture suitable for practical applications. Furthermore, the proposed hyperbolic meta-antennas are directional and show both polarization and angular independence, which is an important property if the meta-antennas are dispersed in solvents or grown on different kind of surfaces.

## II. RESULTS AND DISCUSSION

The first and most important optical property of our system, namely the ability to display in the same platform almost pure radiative (scattering) or non-radiative (absorption) channels at different wavelengths and with the same intensity, is presented in Fig. 1(a), where we plot the absorption (green curve) and scattering (red curve) cross sections, calculated using the finite element method (for more details see Appendix A), of a single hyperbolic meta-antenna with a diameter D = 200 nm, made of 5 alternating layers of gold (10 nm each) and of a dielectric material with n = 1.75 (20 nm each) on a transparent substrate, such as glass (n = 1.5). From now on and where not specified, the environment is considered to be air. We decided to use these dimensions after an optimization study (see Appendix A), to bring this functionality in a specific spectral range, namely the red/near-infrared spectral range (650-1800 nm), which useful for a plenty of emerging light-based technologies and, more importantly, where the constituent multilayered structure displays hyperbolic dispersion of type II (see Appendix A). It is worth mentioning here that, although the structure considered in this case is made of 5 bilayers of gold and dielectric, up to 4 bilayers we can state that our system is still hyperbolic because there are enough bilayers to display the hyperbolic features shown by an infinite multilayer [49, 51]. Furthermore, to better highlight the difference between our architecture and classical plasmonic nanoantennas, we plot also the absorption and scattering cross sections of a gold disk (i) 50 nm thick, namely with the same amount of gold of our meta-anetnna [Fig. 1(b)], and (ii) with the same geometrical characteristics, namely the same diameter and thickness [Fig. 1(c)]. Also these structures are assumed to be on a glass substrate. As it can be inferred by looking at Fig. 1(b)-(c), the gold nanostructures display scattering and absorption at the same wavelength. In this particular case, we can also notice that both architectures show a strong scattering (red curves) and a very low absorption (green curves). This is indeed the expected optical response for plasmonic antennas with these specific sizes and shape in this spectral range. On the contrary, if we look at the hyperbolic meta-antenna we can see two well separated scattering and absorption bands with the same intensity.

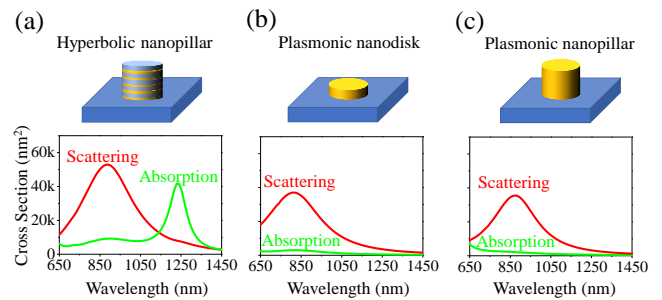

FIG. 1. Scattering (red curves) and absorption (green curves) cross section of (a) a hyperbolic meta-antenna made of 5 bilayers of metal and dielectric (n = 1.75), (b) a classical plasmonic antenna on a glass substrate with the same shape and amount of gold and (c) the same shape and size. All the structures are assumed to stay on a glass substrate.

We want now to go more deeply inside the physical properties of our system. If one wants to control the spectral separation between the scattering and absorption bands, it is actually possible to do so by changing, for instance, the thickness of the metallic or dielectric layers, the shape and size (diameter) of the nanostructure or, in a more convenient way, the dielectric material within our architecture. To demonstrate that the latter possibility can indeed produce a desired and also a significant variation of the spectral separation between the absorption and scattering channel, we chose three different and well-known materials: $SiO_2$ [52], $Al_2O_3$ [53] and $TiO_2$ [54]. The RI of these three dielectrics increases from an average of 1.45 ($SiO_2$) to 2.25 ($TiO_2$). In

Fig. 2(a) we plot the absorption and scattering cross sections of a hyperbolic meta-antenna (D = 200 nm) on a glass substrate made of 5 bilayers of Au (10 nm each) and these three dielectric materials (each layer has a thickness of 20 nm). As can be inferred by Fig. 2(a), the spectral separation between the scattering and the absorption process becomes larger by increasing the RI of the dielectric material. A spectral separation of 250 nm with $SiO_2$ can be more increased up to 580 nm by using $TiO_2$. It is worth noticing that the dependence of the spectral separation between absorption and scattering channels on the value of the RI of the dielectric material chosen is linear, as shown in the inset in the top-panel of Fig. 2(a).

while other two absorption peaks are present at higher wavelength, even if very much smaller in intensity. If we plot the ratio between scattering and absorption, namely $\sigma_{scat}/\sigma_{abs}$, and the inverse of this quantity, we can see, by looking at Fig. 2(b), that these other two absorption peaks. Moreover, it is clear that when the absorption is maximum the scattering is almost totally suppressed, and this is indeed a crucial property if one wants to exploit one or the other effect in the same platform. Moreover, it is worth mentioning here that when the scattering and absorption cross sections are equal, namely $\sigma_{scat}/\sigma_{abs} = 1$ (see the colored lines in Fig. 2(b)), we envision that this system can be used also in plasmon-coupled resonance energy transfer processes at different wavelengths [55]. Furthermore, the three absorption bands highlighted in Fig. 2(b) by using three different colored dots, are related to the excitation of three different localized modes within the meta-antenna, as shown in Fig. 2(c), in analogy with previously reported works where similar confined modes can be excited in a continuous multilayered film, although the latter are guided and not localized [49].

To prove that the spectral separation between scattering and absorption process, as well as its tuning by changing the dielectric material within the structure, is indeed experimentally possible, we fabricated two different samples by keeping as reference the two extreme cases reported in Fig. 2(a), namely hyperbolic meta-antennas on silica substrates made of 5 bilayers of Au and either $SiO_2$ or $TiO_2$. We used a top-down approach based on hole mask colloidal lithography technique [56-58], which is an affordable, highly parallel and $cm^2$-scale nanofabrication method (a detailed explanation of the fabrication process can be found in Appendix B). We controlled the average diameter to be around 200 nm as in the numerical simulations.

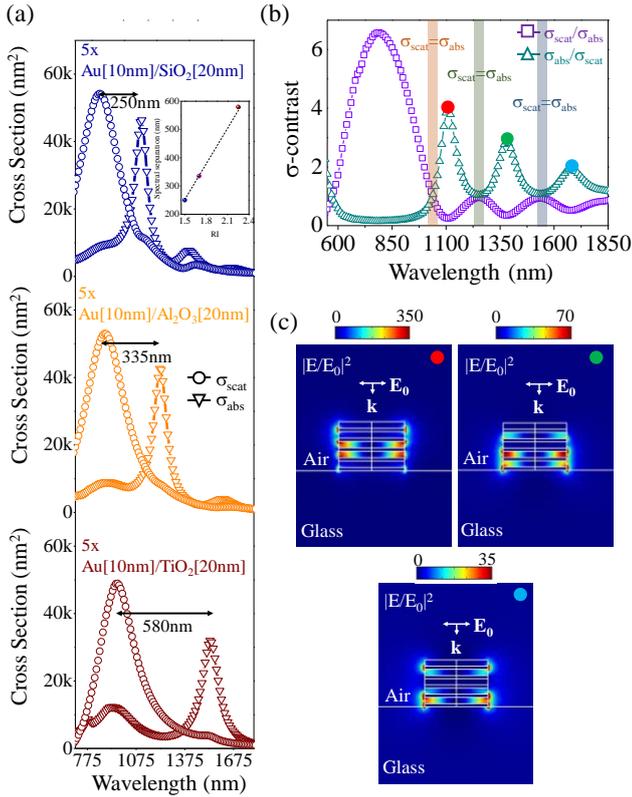

FIG. 2. (a) Calculated scattering (dots) and absorption (triangles) of a hyperbolic meta-antenna with D = 200 nm and made of 5 bilayers of Au (10 nm each) and three different dielectric material ($SiO_2$ – blue curves, top-panel; $Al_2O_3$ – orange curves, middle-panel; $TiO_2$ – red curves, bottom-panel – 20 nm each layer) on a glass substrate. The inset in the top-panel shows the linear dependence of the spectral separation as a function of the RI of the dielectric layers; the dotted line is a guide for eyes. (b) Absorption (cyan triangles) and scattering (violet squares) contrast for the hyperbolic meta-antenna made of 5 bilayers of Au (10 nm each) and $SiO_2$ (20 nm each). (c) Near-field intensity distribution of the three modes highlighted by colored dots in Fig. 2(b).

Moreover, while the scattering peak redshifts less than 100 nm passing from n = 1.45 to n = 2.25, the absorption one displays a redshift of more than 400 nm. If we now focus our attention on the specific case of the Au/$SiO_2$ hyperbolic meta-antenna, it is clear that on the scale of the ordinate of Fig. 2(a) only one absorption peak is clearly visible at λ = 1100 nm,

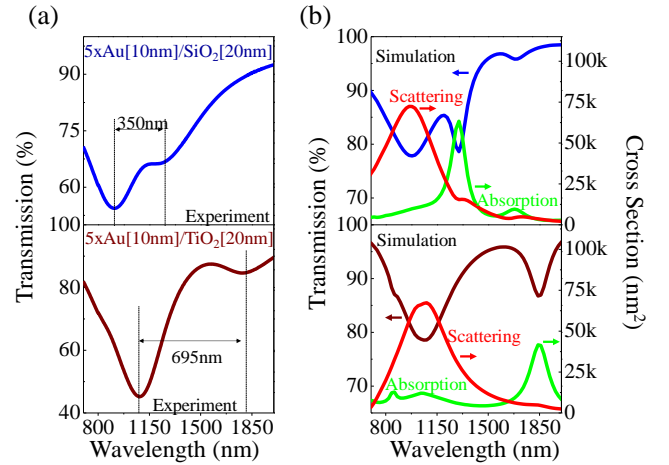

FIG. 3. (a) Measured transmission of hyperbolic meta-antennas on glass (filling factor 20%) with an average diameter D = 200 nm, made of 5 bilayers of Au (10 nm each) and two different dielectric material ($SiO_2$ – blue curve, top-panel; $TiO_2$ – brown curve, bottom-panel). (b) Calculated transmissions and absorption (green curves) and scattering (red curves) cross sections.

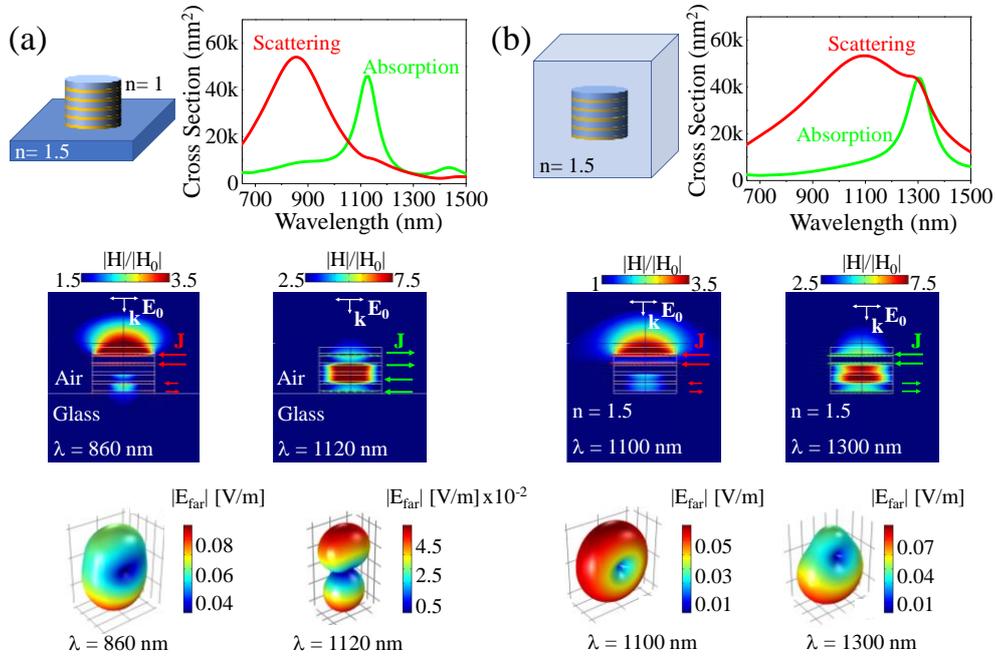

FIG. 4. (a) Top-panel: sketch of a hyperbolic meta-antenna with D = 250 nm and made of 5 bilayers Au (10 nm each) and a dielectric material with n = 1.5 (20 nm each) on a glass substrate (left-panel), and related scattering and absorption cross section as a function of the wavelength of the incoming light (right-panel). Middle panel: magnetic near-field distribution at 860 nm (left panel) and at 1120 nm (right panel). The red and green arrows indicate the direction of the current density J inside the metallic layers. Bottom panel: electric far-field distribution at 860 nm (left panel) and at 1120 nm (right panel). (b) The same as in (a) but for a meta-antenna embedded in a homogeneous medium with n = 1.5.

We then characterized the optical properties of the fabricated samples by measuring their transmission [Fig. 3(a)] (details about the experimental set-up and the optical measurements can be found in Appendix C). In Fig. 3(b) we plot the calculated transmission, which matches almost perfectly the experimental results, including also the calculated scattering (red curves) and absorption (green curves) cross sections. As it can be noticed by looking at the calculated plots, the two transmission dips in the experimental curves can be related either to a pure scattering process or to a pure absorption process. It is worth noticing here that we are able to induced experimentally two well separated decay channels by direct coupling with the far field radiation, which is a very important result also in view of practical applications. Furthermore, it is clear from both the experimental and the calculated curves that by increasing the RI of the dielectric we can increase the separation between the absorption and scattering bands. It is important to mention here that the spectral separation is higher in the experimental case (350 nm and 695 nm for the Au/$SiO_2$-based and Au/$TiO_2$-based meta-antennas, respectively) if compared to the theoretical one in Fig. 2(a), since the experimental effective RI of the $SiO_2$ and the $TiO_2$ layers is a bit higher than the calculated one due to the presence of almost 1 nm of Ti as adhesion layer between each Au and dielectric layer. Moreover, it is also worth noticing that the smaller intensity, compared to that shown in Fig. 2(a) and in Fig. 3(b), of the absorption peak in both the cases studied here can be inferred to several reasons, such as the morphological defects of the multilayers and the presence of the aforementioned Ti adhesion layers, as well as to roughness, round edges, and distribution in size and shape. All these factors can contribute to an increase of the overall losses, in particular at the resonance at 925 nm for the Au/$SiO_2$ sample and at 1095 nm for the Au/$TiO_2$ sample.

Based on this simple proof-of-concept experiment, which proves the robustness of the model, we then used the validated computational approach to understand the main physical mechanisms underlying the formation of almost pure scattering and absorption bands in our system. In Fig. 4(a) we consider a hyperbolic meta-antenna on a glass substrate, with D = 200 nm and made of 5 bilayers of Au (10 nm each) and a dielectric material with n = 1.5 (20 nm each). On the top-right panel of Fig. 4(a) scattering and absorption cross sections as a function of the wavelength are plotted. If we look at the magnetic near-field and at the current density J distributions at the two resonances wavelengths (see middle-panel Fig. 4(a)), we can clearly see a huge difference between the two cases. While at 860 nm (scattering band resonant peak) we have a large contribution from the top of the meta-antenna (first two metallic layers) where the currents within the metal has the same direction, at 1120 nm (absorption band resonant peak) we have that between the first two metallic layers and the last two metallic layers the currents have opposite direction. In the first case we can observe that the magnetic near-field is localized almost outside the meta-antenna, giving rise to the usual far-field pattern of a plasmonic nanoantenna due to the excitation of an electric dipole (bottom-left panel in Fig. 4(a)). In the second case the magnetic near-field is strongly concentrated at the center of the antenna, and the far-field distribution is almost two orders of magnitude lower than that at 860 nm, and this explains the almost total suppression of scattering. Indeed, at λ = 1120 nm we do not observe any scattering peak, but just a huge absorption. This non-radiative coupling between the far-field radiation and the meta-antenna can be reconducted to the excitation of a magnetic dipole, in analogy with previously reported similar effects in metal-insulator-metal (MIM) nanoantennas [59]. It is important here also to notice that the manipulation of plasmonic modes related to electric and magnetic dipoles has been demonstrated to be possible by creating Fano interference between several MIM antennas arranged in a complex fashion [60, 61]. Nevertheless, none of the aforementioned approaches can induce a clear separation between absorption and scattering channel, while in our case the effect is achieved through a straightforward and clear physical concept that avoids the need of sophisticated engineering of the antennas. We want to stress the fact that in our case we need just one single antenna to induce a significant spectral separation between radiative and non-radiative decay channels.

Finally, it is important to mention here that, to display two well-separated bands of almost pure scattering and absorption, it is crucial to have also an index mismatch between the dielectric material in the hyperbolic meta-antenna and the external environment. Due to the strong index contrast, at the magnetic dipole-induced resonance (absorption peak) the electric and magnetic fields are almost totally localized within the meta-antenna, giving rise to strong absorption and to negligible scattering. On the contrary, if we assume that our system is immersed in a homogenous medium with the same index as the dielectric material (in this case n = 1.5, see also the sketch on the top-left panel of Fig. 4(b)), we can observe a huge increasing of the scattering at the magnetic dipole-induced resonance (see the peak of the red curve at 1300 nm in the top-right panel of Fig. 4(b)). In this case we can observe that at 1300 nm the magnetic near-field (middle-right panel of Fig. 4(b)), is still concentrated inside the meta-antenna but there is also a not negligible component outside it. In this case there is not a strong index mismatch between the environment and the dielectric composing the meta-antenna. The index matching between the dielectric composing the nanostructure and the external environment induces a strong quenching of the currents in the bottom side of the meta-antenna. Indeed, we can observe a strong far-field emission (see the bottom-right panel in Fig. 4(b)), which has the same intensity of the electric dipole resonance-induced far-field pattern plotted in the bottom-left panel of Fig. 4(b).

## III. CONCLUSIONS

In summary we have introduced a novel functionality of hyperbolic nanostructured metamaterials. Our proposed architecture displays two well separated scattering and absorption bands. This behavior is related to the excitation of an electric and magnetic dipole, respectively, within the nanostructure, which can be effectively excited by direct coupling with the far field radiation, even when the radiative channel (scattering) is almost totally suppressed, hence making the proposed architecture suitable for practical applications. Furthermore, we have shown that the control of scattering and absorption channels can be achieved over a broad spectral range also by changing the dielectric material within the meta-antennas. Finally, the hyperbolic meta-antennas possess both angular and polarization independent structural integrity (see Appendix A), thus opening up new perspectives for applications on a broad range of surfaces or dissolved in solvents. We foresee that the concept presented here can be generalized by exploring more complex shapes and/or configurations (such as lattice-like configurations) to induce additional or different modes (plasmonic or diffractive) beyond the dipolar modes responsible for the effects shown in this work. The presented findings open the pathway towards novel routes to control the decay channels in light-matter coupling processes beyond what is offered by current plasmon-based architectures, possibly enabling applications spanning, for instance, from thermal emission manipulation, theranostic nano-devices, optical trapping and nano-manipulation, non-linear optical properties, plasmon-enhanced molecular spectroscopy, photovoltaics and solar-water treatments.

## ACKNOWLEDGEMENTS

We acknowledge Matteo Barelli, Andrea Toma and Cristian Ciracì for fruitful discussions.

## APPENDIX A: NUMERICAL SIMULATIONS AND THEORETICAL ANALYSIS

### 1. Optimization of the hyperbolic meta-antennas dimensions and composition

Numerical simulations have been performed using the finite element method implemented in Comsol Multiphysics. The RI values of gold and dielectrics have been taken from literature [52-54, 62]. To simulate the optical properties of hyperbolic meta-antennas we have considered a simulation region where we specified the background electric field (a linearly polarized plane wave), and then we calculated the scattered field by a single meta-antenna to extract optical parameters which are not directly measurable in our laboratory, namely absorption and scattering cross sections. The model computes the scattering, absorption and extinction cross-sections of the particle on the substrate. The scattering cross-section is defined as

$$\sigma_{scat} = \frac{1}{I_0} \iint (\boldsymbol{n} \cdot \boldsymbol{S}) \, dS$$

where $I_0$ is the intensity of the incident light, $\boldsymbol{n}$ is the normal vector pointing outwards from the nanodot and $\boldsymbol{S}$ is the Poynting vector. The integral is taken over the closed surface of the meta-antenna. The absorption cross section equals

$$\sigma_{abs} = \frac{1}{I_0} \iiint Q \, dV$$

where $Q$ is the power loss density of the system and the integral is taken over the volume of the meta-antenna. The transmission is then calculated as

$$T = e^{-\frac{\sigma_{ext} f t}{V}}$$

where $\sigma_{ext} = \sigma_{abs} + \sigma_{scat}$ is the extinction cross section, $f$ is the filling factor (around 20% for the samples fabricated), and $t$ and $V$ are the thickness and the volume of the meta-antenna, respectively.

We have performed an optimization study to find the best configuration, in terms of layers thicknesses and meta-antenna diameter, to maximize the spectral separation between scattering and absorption channels.

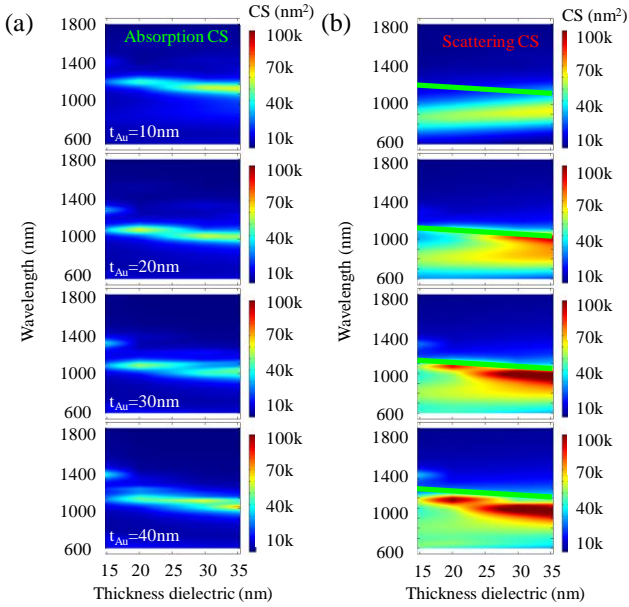

FIG. 5. (a) Absorption and (b) scattering cross sections of a hyperbolic meta-antenna on a glass substrate, with diameter 200 nm and made of 5 bilayers of Au and a dielectric with n = 1.75 as a function of the dielectric thickness, which varies from 10 nm to 40 nm, for Au thicknesses varying from 10 nm to 40 nm from the top to the bottom. The green line in Fig. 5(b) represent the spectral position of the maximum of the absorption cross section.

In Fig. 5(a) we plot the absorption cross section for 4 different cases: from the top to the bottom we plot the absorption cross section for Au thickness ranging from 10 nm to 40 nm, as a function of the wavelength of the incoming light and of the dielectric layer (n = 1.75) thickness, whose range is comprised between 10 nm and 40 nm. In Fig. 5(b) we plot the scattering cross section, indicating with a green line the position of the absorption band resonant peak. The best configuration has been found to be that where the gold thickness is 10 nm and the dielectric thickness is 20 nm.

We have also studied what is the dependence of scattering and absorption cross sections on the meta-antenna diameter. In Fig. 6 we plot the absorption (top-panel) and scattering (bottom-panel) cross sections as a function of the wavelength of the incoming radiation and of the antenna radius. Au and dielectric (n = 1.75) are assumed to have a thickness of 10 nm and 20 nm, respectively. For diameters below 300 nm we have a clear separation between scattering and absorption bands.

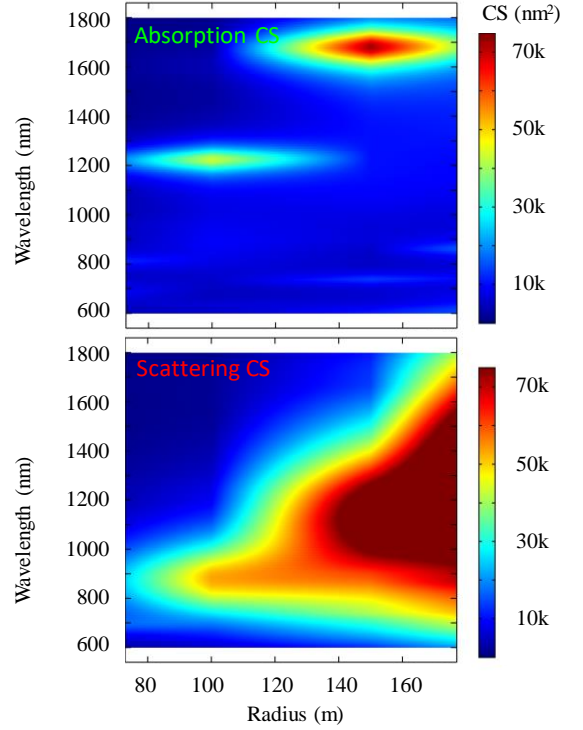

FIG. 6. Absorption (top) and scattering (bottom) cross sections of a hyperbolic meta-antenna on a glass substrate, made of 5 bilayers of Au (10 nm each) and a dielectric (20 nm each) with n = 1.75 as a function of the meta-antenna radius, which is varied from 75 nm to 200 nm.

### 2. Calculation of the effective dielectric constant of a bulk HMM

The dielectric constant of a bulk HMM made of alternating layers of Au and dielectric material with n = 1.75 was calculated using an effective medium approximation. The effective dielectric constant for the multilayered bulk HMM along the two principal directions, namely the x-y plane and z-direction, is calculated as follows [22]

$$\varepsilon_{x,y} = \frac{t_m \varepsilon_m + t_d \varepsilon_d}{t_m + t_d}$$

$$\varepsilon_z = \frac{\varepsilon_m \varepsilon_d (t_m + t_d)}{t_d \varepsilon_m + t_m \varepsilon_d}$$

where $t_m$ and $t_d$ are the thicknesses of Au and the dieletric, respectively, and $\varepsilon_m$ is the dielectric constant of Au, while $\varepsilon_d$ is a dielectric constant of the dielectric layer. In Fig. 7 we plot, as a function of the wavelength of the incident light, the real part of the in-plane (x, y directions, blue curve) and out-of-plane (z direction, violet curve) components of the dielectric tensor.

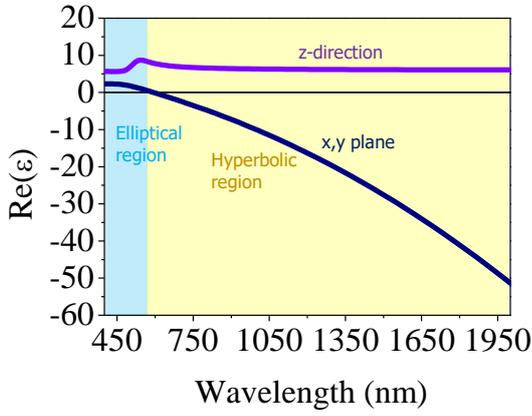

FIG. 7. Real part of the dielectric function of a HMM of type II made of multilayers of Au (10 nm each) and of a dielectric material (20 nm each) with n = 1.75 calculated using the effective medium theory (z-direction component – violet curve; x-y plane component – blue curve.

### 3. Dependence of the optical properties of the hyperbolic meta-antennas on the number of the metal/dielectric layers

A desired control of both $\sigma_{scat}/\sigma_{abs}$ and $\sigma_{abs}/\sigma_{scat}$ is possible by changing the number of the Au/dielectric layers in the meta-antenna. We studied also how the near- and far-field optical distributions change as a function of such a number of layers. In the middle-panel of Fig. 8 we plot the calculated absorption cross section for 8 cases, starting from one bilayer (plasmonic gold nanoantenna 10 nm thick with a capping layer of dielectric with n = 1.75 and 20 nm thick) up to 8 bilayers. All the antennas have a diameter of 200 nm are on glass substrate. As it can be inferred from both the near-field plots (top-panel of Fig. 8), which were calculated at the resonant wavelengths, i.e. where the absorption is maximum, and the absorption curves, up to 3 bilayers – antenna, MIM and metal-insulator-metal-insulator-metal (MIMIM) structures – we do not observe any absorption peak in addition to the one due to the electric dipole-induced LSPR excited throughout the structure. Starting from the 4 bilayers case we can observe at larger wavelengths (around 1300 nm) the rising of an almost pure absorption band due to the excitation of a magnetic dipole within the meta-antenna. It is important noticing that at this wavelength (i) a strong near-field confinement and intensity enhancement (> 300, a factor 3 with respect to the MIMIM case) is observed at the wavelength where the magnetic dipole-induced absorption band is maximum and (ii) the main component of the electric field is is $E_z$ (parallel to the wave-vector **k**) [Fig. 9]. Moreover, as it can be inferred from the central cases of 4, 5 and 6 bilayers, we can actually control either the spectral position where $\sigma_{scat} = \sigma_{abs}$ or the relative intensity between the two maxima of $\sigma_{scat}/\sigma_{abs}$ and $\sigma_{abs}/\sigma_{scat}$ by changing the number of layers. We can pass from a situation where $\sigma_{abs}/\sigma_{scat}$ is higher than $\sigma_{scat}/\sigma_{abs}$ (4 bilayers case) to a case where they are equal (5 bilayers case), to finally arrive to a case where $\sigma_{abs}/\sigma_{scat} < \sigma_{scat}/\sigma_{abs}$. As can be seen from Fig. 8, the best configuration to obtain two distinct and totally de-coupled scattering and absorption bands, namely an almost pure scattering and an almost pure absorption bands with the same efficiency/intensity, is the configuration with 5 bilayers, that is the architecture presented in this work. If we have 6 or more bilayers we start to see a second peak in the scattering cross section at the same wavelength of the magnetic dipole-induced absorption peak, giving rise to a quenching of the absorption contrast $\sigma_{abs}/\sigma_{scat}$, which becomes very much smaller than the scattering contrast $\sigma_{scat}/\sigma_{abs}$ for $n_{bi-layers} > 5$. The second scattering band appearing already in the 6 bilayers case reaches almost the same intensity of the magnetic dipole-induced absorption band once we reach the 8 bilayers case, resembling the response of a 5-bilayers meta-antenna in an index matching case shown in Fig. 4(b). Regarding the main scattering peak present in all the systems considered, from the classical planar plasmonic antenna up to the 8 bilayers hyperbolic meta-antenna, the far-field pattern is the one we expect from a dipolar antenna, as it is actually related to the excitation of an electric dipolar mode. Finally, for completeness, we plot also the far-field distributions of the 6, 7 and 8 bilayers cases at the hyperbolic absorption resonant wavelength, viz. where $\sigma_{abs}/\sigma_{scat}$ is maximum [Fig. 10]. In this case we can observe a strong forward and backward scattering, in contract with the more uniform far-field distribution observed at the electric dipole-induced scattering peak.

### 4. Study of the angular and polarization dependence of the optical properties of the hyperbolic meta-antennas

It is worth mentioning that the optical response of our structure is also strongly independent on both the polarization and the impinging direction of the incident light. In Fig. 11 we plot the absorption and scattering cross sections (top and bottom panels, respectively) of a hyperbolic meta-antenna with D = 200 nm and 5 bilayers of gold and a dielectric with n = 1.75 (each bilayer is composed by 10 nm and 20 nm of material, respectively) as a function of both the wavelength of the incident light and the angle of incidence, for two types of incident waves. More in detail, we consider two linearly polarized plane waves – a transverse magnetic (TM or p-polarized) and a transverse electric (TE or s-polarized) incident field. Up to 70° both the absorption and scattering processes show neither any angular dependence nor any polarization dependence a part for a decreasing of the intensity. Above 60° of incidence we start to see a drop of the scattering and absorption intensities, since the in-plane dipolar LSPR of the metal disks in the meta-antenna are excited with lower efficiency, as the in-plane component of the incident electric field goes to 0. If one wants to take into account also angles larger than 60° or consider, even more in general, a random orientation of the meta-antenna in a homogenous medium, it can be shown that the overall response is that shown for a meta-antenna on a glass substrate with a clear distinction between the scattering and absorption contrast [Fig. 12].

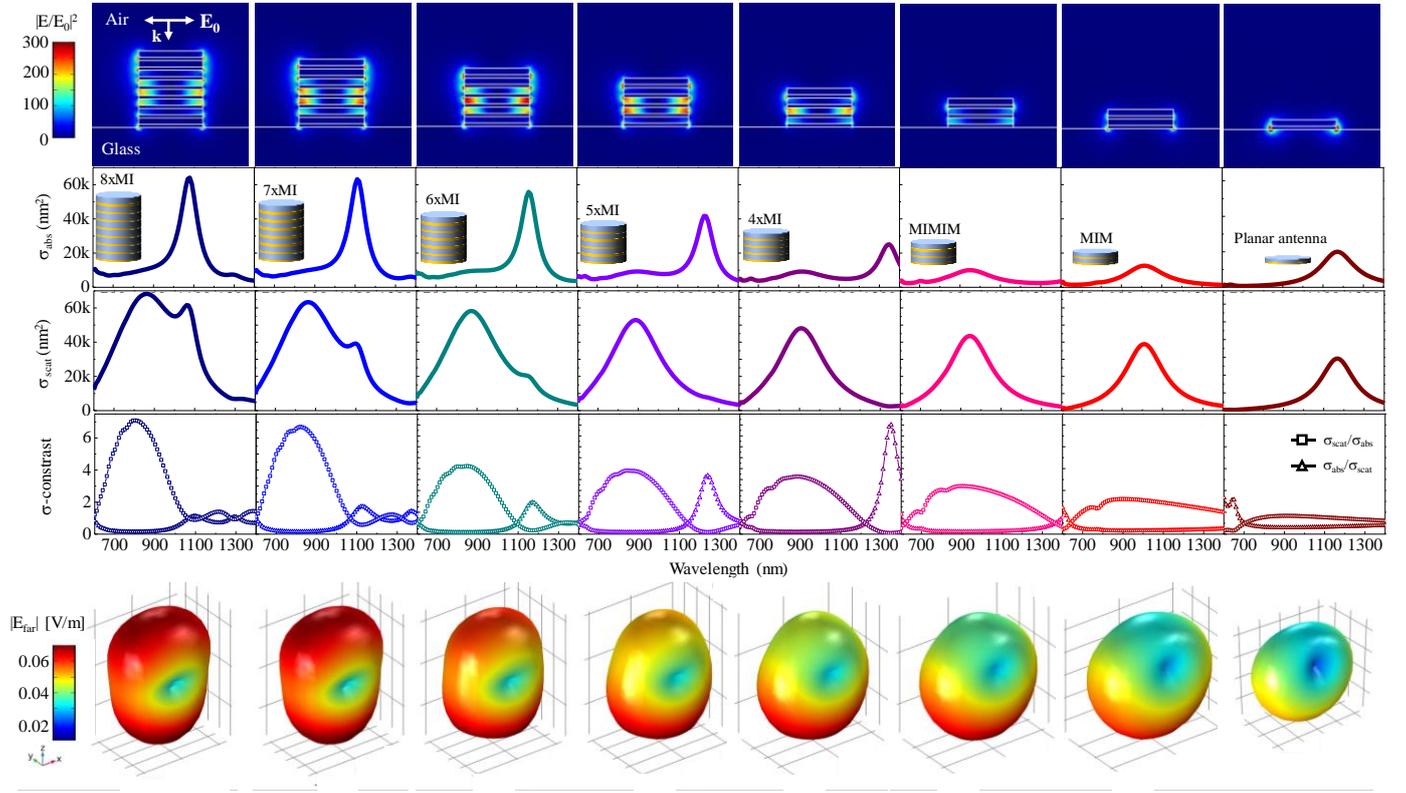

FIG. 8. Top-panel: Near-field intensity distribution of a hyperbolic meta-antenna on a glass substrate in air at normal incidence and at the resonant wavelength of the magnetic dipole-induced absorption band. Middle-panel: absorption (top), scattering (middle) and A and S (bottom panel) evolution upon variation of the number of bilayers. Bottom-panel: far-field distribution at the resonant wavelength of the electric dipole-induced scattering band.

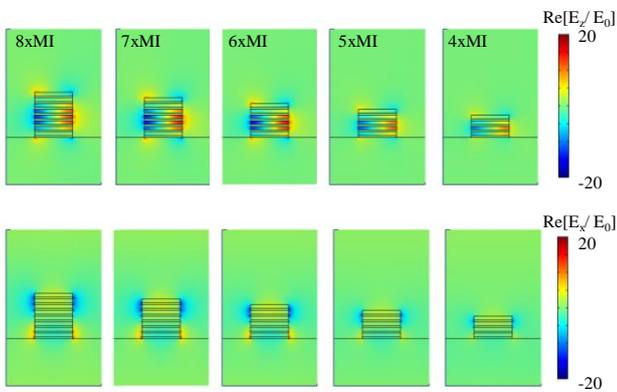

FIG. 9. Near-field intensity distribution of the z- (top panel) and x-component (bottom panel) normalized to the incident plane wave intensity $E_0$ for the cases 4-8 bilayers of Fig. 8.

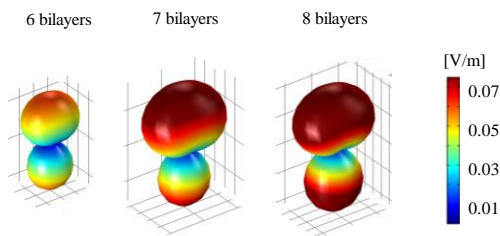

FIG. 10. Far-field distribution at the wavelength of the magnetic dipole-induced absorption peak for the cases of 6, 7 and 8 bilayers of Fig. 8.

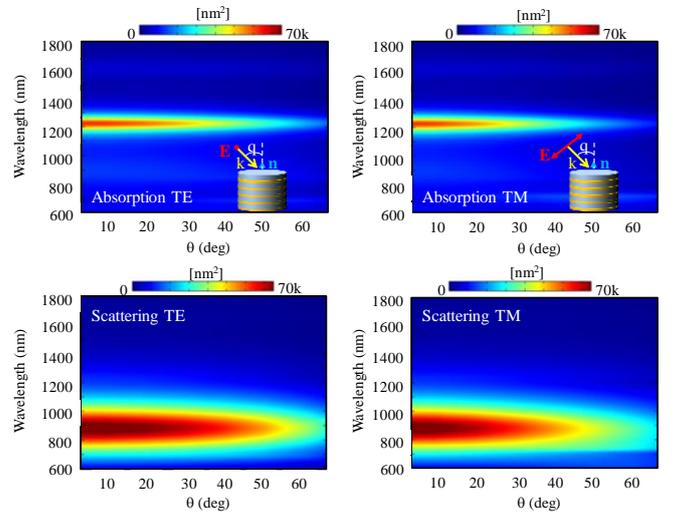

FIG. 11. Top-panels: absorption cross section for TE (left-panel) and TM (right-panel) polarization of the incident light as function of the wavelength and of the angle of incidence. Bottom-panels: scattering cross section for TE (left-panel) and TM (right-panel) polarization of the incident light as function of the wavelength and of the angle of incidence.

The polarization/angular independence of our architecture is of great significance since it means that the special optical property of the system reported here does not depend on the orientation of the meta-antenna. This fact implies that our system can be randomly deposited on different surfaces and implemented in a large range of practical applications, for instance in plasmon-based photovoltaic devices [63] or solar transparent radiators [64], as well as they can be dissolved in solution for biomedical applications [65].

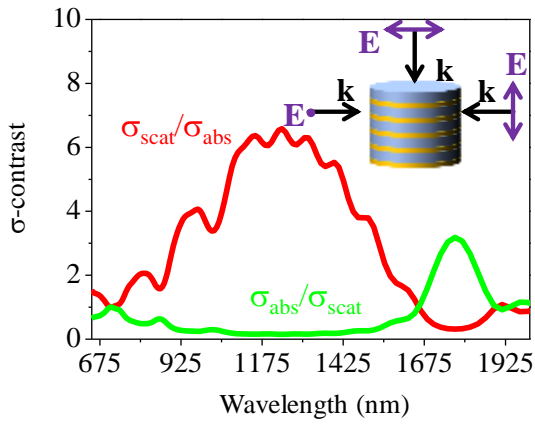

FIG. 12. Calculated $\sigma_{scat}/\sigma_{abs}$ (red curve) and $\sigma_{abs}/\sigma_{scat}$ (green curve) of a randomly oriented hyperbolic meta-antenna made of 5 bilayers of Au (10 nm each) and dielectric with n = 1.75 (20 nm each) in water (n = 1.33). The curves are obtained by making a convolution of the cross sections obtained for the three different orientations between the incident electric field and the meta-antenna as shown in the inset.

**APPENDIX B: SAMPLE FABRICATION**

Hyperbolic meta-antennas were prepared by inductively coupled plasma (ICP) etching of the gold/dielectric multilayers with the Cr disk as mask, which was fabricated by hole mask colloidal lithography [56-58]. With this approach it is possible to fabricate large areas of hyperbolic meta-antennas with the predicted properties, which can be easily transferred on other substrates or disperse in solution, as demonstrated recently by some groups who already proposed detailed and efficient protocols [66, 67, 64].

**1. Stacking bi-layers fabrication.**

Microscope glass slides were cleaned with acetone and 2-propanol with 2min sonication respectively. After deionized water (DI) washing and blow drying under $N_2$ flow, the glass wafers were ready for the multilayer deposition. For the Au/$SiO_2$ stacking layer deposition, the glass wafers were loaded into an electron beam deposition (E-beam, PVD75 Kurt J. Lesker company) chamber. One unit of the metal-dielectric bi-layer consisted of 0.5nm Ti +10 nm Au + 0.5 nm Ti + 20 nm $SiO_2$, in which Ti served as the adhesion layer. The deposition of the bi-layer unit was repeated five times. For the Au/$TiO_2$ stacking layers, the glass wafers were loaded into an electron beam deposition chamber (Kenosistec KE 500 ET), and 0.5 nm Ti + 10 nm Au + 0.5 nm Ti layers were deposited at a rate of 0.3 Å/s. The wafer was then transferred to an atomic layer deposition chamber (ALD, FlexAL, Oxford Instruments) and $TiO_2$ was deposited using a process with titanium isopropoxide as the titanium precursor and oxygen plasma as the oxidizer. The process was repeated at 80 °C temperature for 383 cycles to produce a film with a thickness of 20nm, which was verified with ellipsometry.

One unit of the Au/$TiO_2$ metal dielectric bi-layer consisted of 0.5 nm Ti + 10 nm Au + 0.5 nm Ti + 20 nm $TiO_2$. The deposition of the bi-layer unit was repeated five times.

**2. Cr disk etching mask fabrication.**

On the top of stacking bilayers, photoresist (950 PMMA A8, Micro Chem) was spin coated at 6000 rpm and soft baked at 180 °C for 1min. After $O_2$ plasma treatment (2min, 100W, Plasma cleaner, Gambetti), Poly(diallyldimethylammonium chloride) solution (PDDA, Mw 200,000-350,000, 20 wt. % in H2O, Sigma, three times diluted) was drop coated on the top of the PR surface and incubated for 5min to create a positively charged surface. The extra PDDA solution was washed away under flowing DI water after 5min incubation. Then negatively charged polystyrene(PS) beads (diameter 552nm, 5wt% water suspension, Micro Particle GmbH ) were drop coated on the as prepared stacking bi-layers, cleaned after 30s under flowing DI water and dried with $N_2$ flow. Thereby, random distributed PS beads were attached on top of the photoresist. The samples were treated with $O_2$ plasma etching in the inductively coupled plasma-reactive ion etching system (ICP-RIE, SENTECH SI500) to reduce the size of PS beads. Gold film (40nm) was sputter coated (Sputter coater, Quorum, Q150T ES) on top of the sample to serve as an etching mask to protect the PR underneath. After removal of the PS beads by Polydimethylsiloxane (PDMS) film, the samples were treated again by $O_2$ plasma in the ICP-RIE system to etch away the PR and create randomly distributed holes as mask on top of the stacking bi-layers. The diameter of the holes was controlled by varying the PS bead $O_2$ plasma treatment time. E-beam deposition of 100nm Cr was then performed with a vertical incident angle. Followed by liftoff of the PR in acetone, randomly distributed Cr disks on the stacking multilayer were fabricated.

**3. Meta-antennas fabrication.**

With the Cr disk mask, ICP-RIE etching was carried out with $CF_4$ gas flow 15sccm, radio frequency (RF) power 200 W, ICP power 400W, temperature 5°C, pressure 1Pa. The etching time was adjusted according to the stacking film thickness to ensure all the extra stacking bi-layer material, except for the area under Cr mask, was removed. Then the sample was soaked in Cr etchant (Etch 18, Organo Spezial Chemie GmbH) for 2min to remove the Cr mask. Followed by DI water cleaning and drying under $N_2$ flow, the sample morphology was characterized with a scanning electron microscope (SEM). In Fig. 13 we show a representative SEM image of the randomly distributed Au/$SiO_2$ meta-antennas fabricated on a glass substrate. As shown in the inset, the stacking layers can be well distinguished indicating that we were able to fabricate multilayered nanostructures without damaging the multilayers while maintaining the structural integrity.

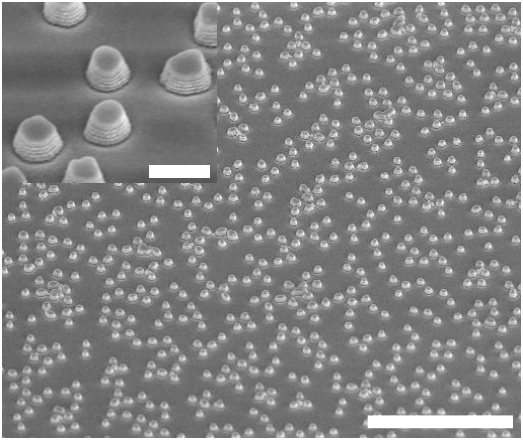

FIG. 13. SEM image (scale bar 4 μm) of hyperbolic meta-antennas fabricated using hole mask colloidal lithography on top of a multilayer of Au and SiO$_2$. Inset image: detail of the fabricated structures showing the multilayered structure of the single meta-antenna (scale bar 400 nm).

## APPENDIX C: OPTICAL CHARACTERIZATION

Cary 5000 UV-vis- near infrared spectrophotometer was used for the measurement of transmission spectra. Before measuring the samples, the baseline correction was performed by collecting the 100% transmission from a glass slide and 0% transmission with blocked probing light. The incident light is unpolarized. The transmission spectra were collected from the samples with the meta-antennas facing the incident light. To collect angle resolved transmission spectra, the samples were rotated from the normal incident position. In Fig. 14 we plot the transmission spectra of the sample made of 5 bi-layers Au/TiO$_2$ at three different angles of incidence: 0°, 30° and 60° to prove the polarization and angular independence predicted by the calculations of absorption and scattering cross sections plotted in Fig. 11.

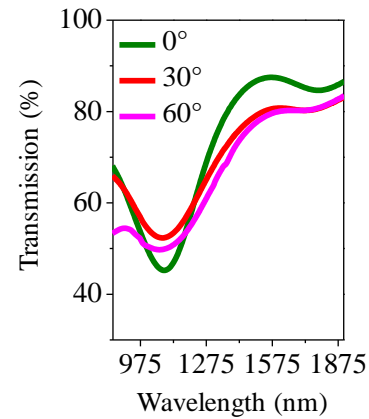

FIG. 14. Experimental transmission of hyperbolic meta-antennas on glass made of 5 bilayers of Au and TiO$_2$ at different angles of incidence, 0°, 30° and 60°. The incident light is unpolarized.